\documentclass[english,submission]{eptcs}
\usepackage[T1]{fontenc}
\usepackage[latin9]{inputenc}
\usepackage{listings}
\usepackage{amstext}
\usepackage{graphicx}

\makeatletter

\pdfpageheight\paperheight
\pdfpagewidth\paperwidth

\usepackage{stmaryrd}
\usepackage{amstext}
\usepackage{amssymb}

\usepackage[protrusion=true,expansion]{microtype}

\usepackage{tikz}
\usetikzlibrary{calc}
\usetikzlibrary{patterns}
\usetikzlibrary{fit}

\usetikzlibrary{shapes.geometric}
\usetikzlibrary{arrows}

\tikzset{module/.style={draw,anchor=north west}}
\tikzset{procedure/.style={draw,anchor=north west}}
\tikzset{situation/.style={draw,thin,rounded corners,inner sep=4pt,outer sep=0pt,minimum size=6mm,anchor=north west}}
\tikzset{precondition/.style={situation,very thick}}
\tikzset{postcondition/.style={situation,double}}
\tikzset{invariant/.style={anchor=north west}}
\tikzset{transition/.style={draw,thick,outer sep=0pt}}
\tikzset{statement/.style={outer sep=0pt,anchor=north west,font=\small}}
\tikzset{branch/.style={inner sep=2pt, outer sep=0pt, shape=circle, fill,thin}}
\tikzset{anch/.style={inner sep=0pt, outer sep=0pt}}
\tikzset{blackdot/.style={inner sep=1.5pt, outer sep=0pt, shape=circle, draw, fill=black,thin}}
\tikzset{if/.style={inner sep=1.75pt, outer sep=0pt, shape=rectangle, fill=black,draw,thin}}
\tikzset{choice/.style={inner sep=1.75pt, outer sep=0pt, shape=rectangle, fill=white,draw,thin}}
\tikzset{whitedot/.style={inner sep=0.2ex, outer sep=0pt, shape=circle, draw, fill=white,thin}}

\newcommand{\procedurelabel}[1]{\small \texttt{#1}}
\newcommand{\situationlabel}[1]{\small \textsc{#1}}

\newcommand{\situationrule}{\vspace{-1.2ex}\mbox{}\rule[1.2ex]{\linewidth}{0.4pt}\mbox{}}

\newcommand{\procedurerule}{\situationrule}

\newcommand{\invariant}[1]{\small \ensuremath{\mathtt{#1}}}

\newcommand{\declaration}[1]{\small \ensuremath{\mathtt{#1}}}

\newcommand{\ibpid}[1]{\texttt{#1}}

\newcommand{\ibpkw}[1]{\textsf{\textbf{#1}}}

\tikzset{elm/.style={draw,anchor=south west,outer sep=0pt},  rng/.style={draw,anchor=south west,outer sep=0pt}, sep/.style={draw,thick,anchor=south west,outer sep=0pt,inner sep=0pt,fill=black}, idx/.style={right,font=\small}, stmt/.style={font=\small}, heap/.style={fill=gray!50}, arr/.style={matrix,anchor=north west,column sep={-0.4pt,between borders}}, lheap/.style={pattern color=gray!70,pattern=north east lines} }

\tikzset{situation/.style={draw,thin,rounded corners,inner sep=4pt,outer sep=0pt,minimum size=6mm,anchor=north west,fill=yellow!10}}

\makeatother

\usepackage{babel}
\begin{document}

\title{An Exercise in Invariant-based Programming with Interactive and Automatic Theorem Prover Support}

\author{Ralph-Johan Back \qquad\qquad Johannes Eriksson
\institute{Department of Information Technologies\\
\r{A}bo Akademi University\\
Turku, Finland}
\email{backrj@abo.fi \qquad\qquad\qquad joheriks@abo.fi} }

\def\titlerunning{An Exercise in IBP with Interactive and Automatic
Theorem Prover Support}
\def\authorrunning{R.-J. Back \& J. Eriksson}

\def\event{THedu'11}

\maketitle

\begin{abstract}
  \emph{Invariant-Based Programming (IBP)} is a diagram-based
  correct-by-construction programming methodology in which the program
  is structured around the invariants, which are additionally
  formulated \emph{before }the actual code. \emph{Socos} is a program
  construction and verification environment built specifically to
  support IBP. The front-end to Socos is a graphical diagram editor,
  allowing the programmer to construct invariant-based programs and
  check their correctness. The back-end component of Socos, the
  \emph{program checker}, computes the verification conditions of the
  program and tries to prove them automatically. It uses the theorem
  prover PVS and the SMT solver Yices to discharge as many of the
  verification conditions as possible without user interaction. In
  this paper, we first describe the Socos environment from a user and
  systems level perspective; we then exemplify the IBP workflow by
  building a verified implementation of heapsort in Socos. The case
  study highlights the role of both automatic and interactive theorem
  proving in three sequential stages of the IBP workflow: developing
  the background theory, formulating the program specification and
  invariants, and proving the correctness of the final implementation.
\end{abstract}

\section{Introduction\label{sec:Introduction}}

\emph{Invariant-based programming }(IBP) is a method for formal
verification of imperative programs \cite{jBack08b}. It is a
\emph{correct-by-construction }method: the correctness proofs are
developed hand-in-hand with the program. In IBP the internal loop
invariants of the program are also written \emph{before} the code.
After the invariant structure has been established, the code is added
in small increments, and each extension is verified to preserve the
invariants. Letting the correctness arguments determine the structure
of the code, rather than vice versa, makes the verification task
significantly less difficult compared to verification \emph{a
  posteriori}. IBP has been successfully applied as a pedagogical
device in teaching introductory formal methods \cite{inpBaErMa07}.

The correctness of even small programs depends on a large number of
\emph{verification conditions} to be proved. We are building a programming
environment called \emph{Socos}%
\footnote{\texttt{http://www.imped.fi/socos}%
}, which applies state-of-the-art automatic theorem proving tools and
satisfiability modulo theories (SMT) solvers to discharge as many
of the lemmas as possible without user intervention. The front-end
to the system is a graphical \emph{diagram editor}, supporting both
constructing the program and checking its correctness. This front-end
is implemented as a plug-in for Eclipse \cite{eclipse}. The back-end
\emph{program checker} derives the verification conditions from the
program source, and interfaces with the theorem prover PVS \cite{owre96pvs}
to automatically discharge as many of the conditions as possible.
Socos allows the full higher-order logic of PVS in specifications
and invariants. Hence, all conditions could not be proved automatically.
Conditions that were not automatically discharged can be proved interactively
in the PVS proof assistant. Alternatively, proof automation can often
be improved by introducing abstractions which are more suitable for
automatic reasoning in the domain of discourse. Such abstractions
can be added to the verification process through \emph{background
theories}, and domain-specific proof strategies based on background
theories can significantly improve proof automation.

This paper presents the workflow of Socos-supported IBP in the context
of a case study. We first describe IBP in general, followed by a overview
of Socos from both a user and a systems level perspective. Next, we
build a set of PVS background theories for dynamic arrays, sortedness
and permutations. Finally, based on these theories we build a verified
implementation of heapsort. The case study focuses on the interplay
between programming and proving, and describes how the complete workflow
from specification to verified implementation is supported by Socos.
Although the code itself is small, verification of heapsort involves
several nontrivial invariants and proofs. The specification involves
the notions of sortedness, permutations, and heaps. We extend the
background theories by proving additional lemmas in PVS to improve
automation while maintaining soundness with respect to the base definitions.
The case study also shows how Socos can identify bugs related to corner
cases, which are otherwise easily missed during testing.

\paragraph{Related work.}

IBP builds on early work by Back \cite{Back78:situation}, Reynolds
\cite{Reynolds78}, van Emden \cite{vEmden79}. A comprehensive overview
of the method is given in \cite{jBack08b}. A description of the semantics
and proof theory of IBP can be found in \cite{back_preoteasa11proofrules}.
There exists a large number of verification tools based on VC generation
and theorem proving. PVS verification of Java programs is supported
by Loop \cite{vandenberg01loop} and the Why/Krakatoa tool suite \cite{citeulike:2868751}.
Several program verifiers are based on SMT solvers. Boogie \cite{barnett_boogie:modular_2006}
is an automatic verifier of BoogiePL, a language intended as a backend
for encoding verification semantics of object oriented languages.
Spec\#, an extension to C\#, is based on Boogie \cite{specSharpOverview}.
Back and Myreen have developed an automatic checker for invariant
diagrams \cite{inpBaMy05a} based on the Simplify validity checker
\cite{1066102}. Together with the second author they later developed
the checker into a prototype of the Socos environment \cite{inpBaErMy07}.

\paragraph{Overview of paper.}

The remainder of the paper is as follows. Section \ref{sec:Invariant-diagrams}
introduces the notion of invariant diagrams and their correctness.
Section \ref{sec:Invariant-based-programming-in-Socos} describes
the Socos environment from the user perspective. Section \ref{sec:System-overview}
gives a systems-level overview of Socos, focusing on the interface
to the underlying components (PVS and Yices). In Section \ref{sec:Background-theories}
we develop a background theory for dynamic arrays, sortedness and
permutations. Section \ref{sec:Example:Heapsort} develops the case
study, a verified implementation of heapsort. Section \ref{sec:Conclusion}
concludes the paper with a summary and some observations.

\section{Invariant diagrams\label{sec:Invariant-diagrams}}

\begin{figure}
\noindent \begin{centering}
\includegraphics[width=1\textwidth]{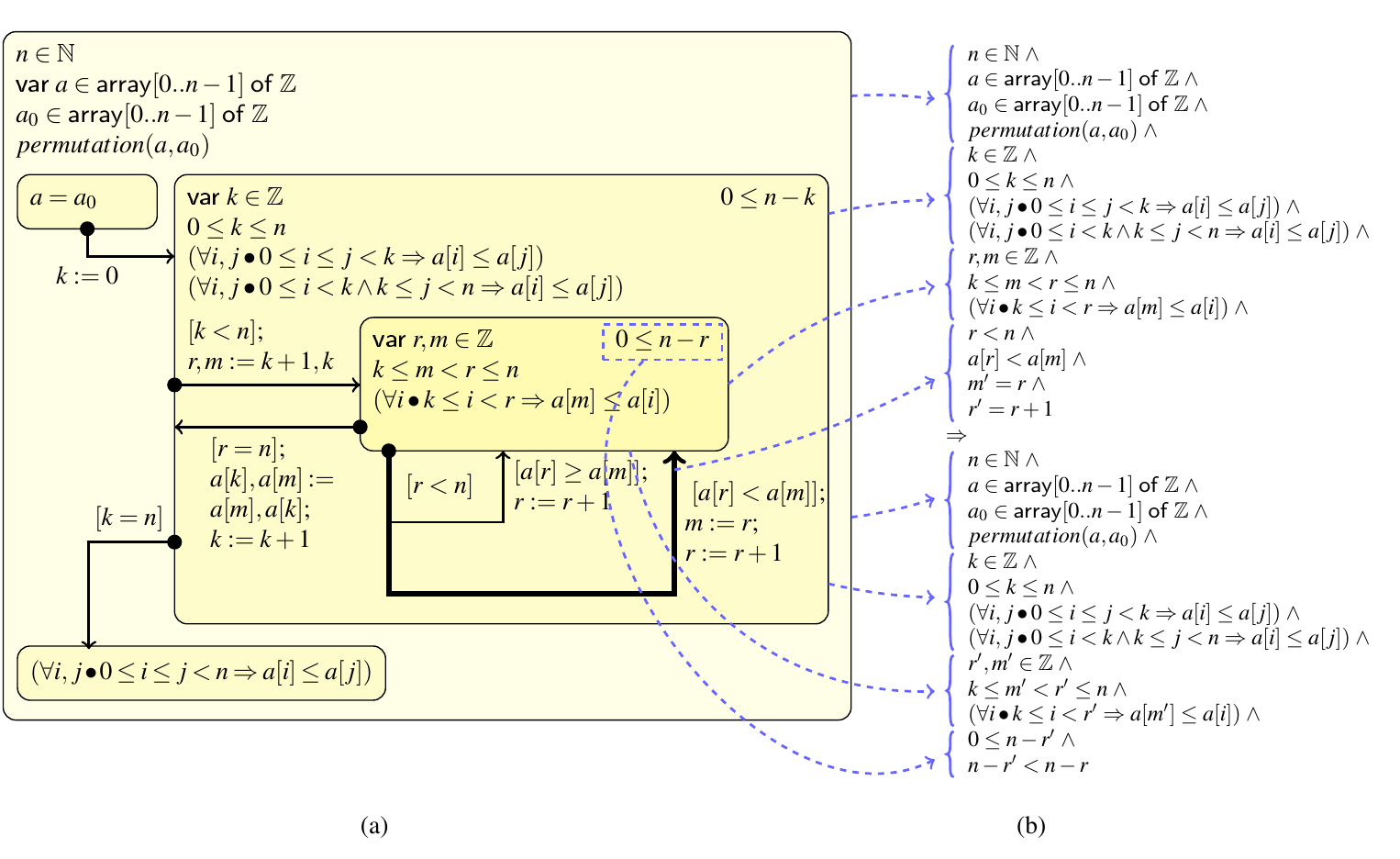}
\par\end{centering}

\caption{(a) invariant diagram for selection sort; (b) consistency and termination
conditions for the bolded loop transition\label{fig:Sorting}}
\end{figure}

The basic building blocks of invariant-based programs are
\emph{situation}s and \emph{transitions.} Situations are predicates
over the state space of the program, whereas transitions are program
statements.  \emph{Invariant diagrams} are directed, nested graphs
where the nodes correspond to situations and the edges correspond to
transitions. The operational interpretation of an invariant diagram is
that of a state chart: control flows from situation to situation by
(nondeterministically) following \emph{enabled }transitions. A
transition is enabled if its \emph{guard }holds in the current state.
Figure \ref{fig:Sorting}a shows an IBP implementation of the
\emph{selection sort} algorithm. Situations are drawn as rectangles
with rounded corners, transitions as arrows connecting the rectangles.
The predicate (invariant) of a situation is written in the top left
corner of the situation. Statements---sequential composition of guards
and assignments---are written adjacent to the transition arrows. The
program consists of an inner and an outer loop. Each iteration of the
outer loop extends the sorted portion with one element by finding (in
the inner loop) the minimal element in the unsorted portion (at index
$m$) and then exchanging it with the first element in the unsorted
portion (at index $k$). The invariant of the inner loop is stronger
than that of the outer loop. Nesting the inner loop situation inside
the outer loop situation indicates that the invariant of the outer
loop should be inherited.

An invariant-based program is \emph{correct} if execution, when
started from any one situation, terminates in a \emph{final
  situation}. A final situation is a situation with no outgoing
transitions. Final situations correspond to the postcondition(s) of the
program. An invariant diagram can be interpreted as a total
correctness theorem, where each transition corresponds to a
\emph{consistency} lemma, each intermediate (non-final) situation
corresponds to a \emph{liveness} lemma, and each loop corresponds to a
\emph{termination} lemma. A transition is consistent if the source
situation, the guard and the assignments imply the target situation.
An intermediate situation is live if at least one outgoing transition
is always enabled. A loop is terminating if each cycle strictly
decreases a \emph{termination function}, i.e., a function from the
program states to a well-founded set.  The termination function is
written together with its lower bound in the upper right hand corner
of the recurring situation. A diagram is correct iff all transitions
are consistent, all intermediate situations are live, and all loops
are terminating.

The programmer first defines the situation structure, and then adds
and checks the transitions one by one. The lemma to be checked for a
transition can be read directly from the diagram. Figure
\ref{fig:Sorting}b shows the condition for the loop transition in the
example. The antecedent contains the source situation predicate, the
guard of the transition, and the equalities introduced by the
assignments to variables $m$ and $r$. The consequent contains the same
situation predicates over the updated values $m'$ and $r'$, and
additionally a constraint that the termination function of the inner
loop ($n-r$) remains bounded from below (by $0$) while strictly
decreasing.

\section{Invariant-based programming in the Socos environment\label{sec:Invariant-based-programming-in-Socos}}

Socos supports construction and static checking of invariant-based
programs. The top level document, called the \emph{verification
context}, defines global constants, associated PVS background
theories, and a default proof strategy. Nested within the verification
context is a collection of (mutually recursive) procedures. Each
procedure is specified by a precondition and one or more
postconditions, and implemented by an invariant diagram. Visually,
pre- and postconditions are distinguishable from intermediate
situations by the outline: preconditions are drawn with a thick
outline, whereas postconditions are drawn with a double outline. If
the precondition is omitted, it defaults to true and the initial
transition is drawn from the procedure outline. The transition
language supports sequential composition of assumptions, assertions,
assignments, and procedure calls. All expressions, including guard
expressions and the right hand side of assignments, are written in the
PVS syntax.

\begin{figure}[h]
\begin{centering}
\includegraphics[width=0.95\columnwidth]{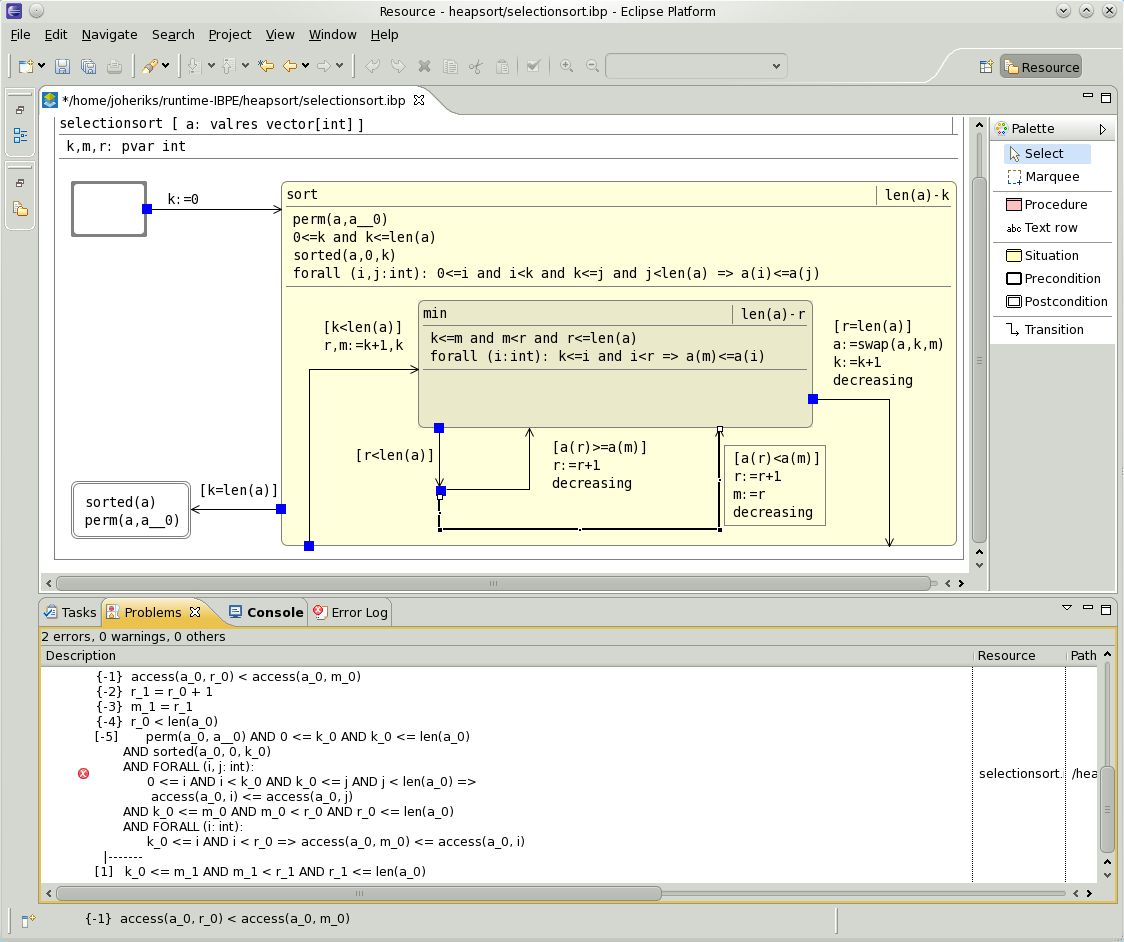}
\par\end{centering}

\caption{The Socos programming environment\label{fig:SelectionSort}}
\end{figure}

The programmer edits the verification context and its contained diagrams
in a graphical environment (Figure \ref{fig:SelectionSort}). By the
click of a button, Socos generates the verification conditions from
the diagram, attempts to discharge as many as possible automatically,
and then reports the unproved conditions to the programmer. 
Figure \ref{fig:SelectionSort} shows a session in which the
program in Figure \ref{fig:Sorting}, implemented as a Socos procedure,
is being checked. In this case, the program contains an error: the
second loop transition has the increment $\mathtt{r:=r+1}$ and assignment
$\mathtt{m:=r}$ in the wrong order. Consequently, the loop invariant
is not preserved by the transition. Socos pinpoints the inconsistency
by highlighting the loop transition, and the unproved (false) condition
associated with the transition becomes visible in the {}``Problems
view''. 

Invariant diagrams are built and checked incrementally, i.e., transition
by transition. Hence, all transitions may not be in place when the
program is checked. Consistency is always checked for all transitions
that have been added so far to the diagram. Liveness and termination
checking can be postponed. For instance, omitting the termination
function disables generation of termination conditions, and instead
Socos prints a warning that the program may not be terminating.

\section{System overview\label{sec:System-overview}}

\begin{figure}[h]
\begin{centering}
\includegraphics{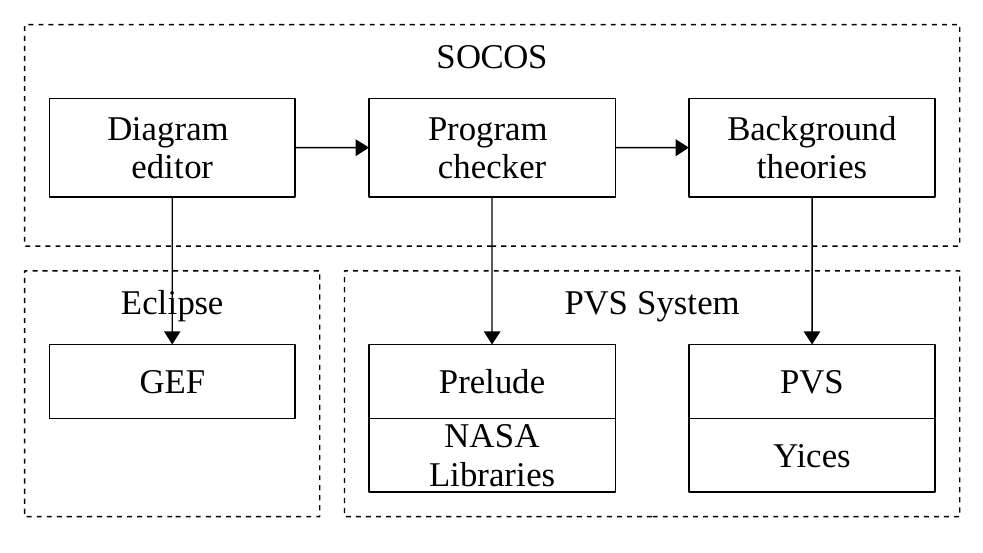}
\par\end{centering}

\caption{Software architecture\label{fig:Architecture}}
\end{figure}

Figure \ref{fig:Architecture} shows the components of Socos and their
interdependencies. In this section, we briefly describe these components.

\subsection{Diagram editor}

The diagram editor is implemented as an extension to Eclipse
\cite{eclipse}, an extensible platform for tool integration. Eclipse
extensions, called \emph{plug-ins}, implement a set of standardized
extension points provided by Eclipse to implement the functionality of
the plug-in.  The user interface of Eclipse follows a workspace
metaphor, in which the user manages a set of resources through
\emph{views }and \emph{editors}.  A view is a UI component displaying
a resource; editors allow both viewing and updating a resource.  The
Socos plug-in adds an invariant diagram editor built on top of the
\emph{Graphical Editing Framework (GEF)} provided by Eclipse. The
editor's associated tool palette, shown in the right hand side of
Figure \ref{fig:SelectionSort}, contains tools for code editing,
situation placement, and transition routing. Clicking the {}``check
button'' sends the diagram to the program checker, which can be
called either locally (over Unix pipes) or remotely (over http).

\subsection{Program checker}

The program checker generates a PVS translation of the verification
conditions for the diagram. The verification conditions are
calculated by weakest preconditions, and exported into a PVS theory
file containing a lemma for each condition. To each lemma, the program
checker also associates a proof script which is run through the PVS
proof checker, and the final proof state (proved or failed) of each
condition is collected. Any PVS strategy can be used to attempt to
discharge the conditions; the default strategy invokes Yices. We give
here only a brief overview of the underlying proof tools and the translation;
the verification semantics is described in detail in \cite{inpErBa10}.

\paragraph{PVS and Yices.}

PVS%
\footnote{\texttt{http://pvs.csl.sri.com}%
} is a free, open source theorem proving system based on simply-typed
higher-order logic \cite{PVS-Semantics:TR}. It provides base types
such as \texttt{bool}, \texttt{nat}, \texttt{int} and \texttt{real},
and type constructors to build new types from existing types. Types
are related to sets: two types are equal if they denote the same set
of values, and subtypes correspond to subsets. For example, $\mathtt{nat}$
is a subtype of $\mathtt{int}$, $\mathtt{int}$ is subtype of $\mathtt{rational}$,
and $\mathtt{rational}$ is a subtype of $\mathtt{real}$. Subtypes
are introduced by \emph{predicate subtyping} \cite{Rushby98:TSE};
the subtype is defined by a predicate on the supertype. Type checking
in PVS is undecidable in the general; \emph{type correctness conditions
}(TCCs) generated by the type checker may hence require interactive
proof.

PVS proof theory is based on \emph{sequent calculus}. A proof is a
tree where each node is a sequent of the form $\gamma_{1},\dots,\gamma_{n}\vdash\delta_{1},\dots,\delta_{m}$
where $\gamma_{1},\dots,\gamma_{n}$ are the antecedents and
$\delta_{1},\dots,\delta_{m}$ are the consequents. PVS proofs are
goal-directed: the proof of a proposition $\alpha$ starts with the
root sequent $\vdash\alpha$. A command either proves a sequent, or
reduces it to subgoals. A proof tree is complete when every leaf is
proved. The logic of PVS is embodied in a small set of primitive inference
rules. Every command corresponds to a sequence of applications of
these rules. Proof \emph{strategies} are higher-order functions combining
basic commands into more powerful commands.

Yices%
\footnote{\texttt{http://yices.csl.sri.com}%
} is a free SMT solver which can be used as a decision procedure in
PVS \cite{yices}. To check the validity of a sequent $\gamma_{1},\dots,\gamma_{n}\vdash\delta_{1},\dots,\delta_{m}$,
the command \texttt{$\texttt{(yices)}$} checks the satisfiability
of the formula $\gamma_{1}\land\dots\land\gamma_{n}\land\neg\delta_{1}\land\dots\land\neg\delta_{m}$
using Yices. If the formula is unsatisfiable, the sequent is valid
and is thus discharged; otherwise, \texttt{$\texttt{(yices)}$} does
nothing.

\paragraph{Verification condition generation.}

The consistency condition for a transition $S_{X,Y}$ from situation
$X$ to situation $Y$ is generated based on the rule: 
\[
\forall\sigma : P_{X}(\sigma)\Rightarrow wp(S_{X,Y})(P_{Y})(\sigma)
\]
The variable $\sigma$ ranges over all program states, $P_{X}$ and
$P_{Y}$ are the state predicates of the situations $X$ and $Y$,
and $wp(S_{X,Y})$ is the weakest precondition predicate transformer
for the statement $S_{X,Y}$. Based on this rule, one PVS lemma is
generated for each situation, capturing the consistency of all outgoing
transitions. Procedure calls are verified consistent based on the
pre- and postconditions of the called procedure in the usual way.

A procedure is live if the following conditions both hold: (1) the
postcondition is reachable from the precondition; and (2) each statement
can proceed from any state it may be reached by (absence of miracles).
Condition (1) is checked by analyzing the transition graph. Condition
(2) is true for all statements satisfying the {}``excluded miracle''
law: $\forall\sigma : \neg\text{wp}(S)(\emptyset)(\sigma)$. Assignments,
procedure calls and guarded choices satisfy this property. Socos
also allows assume statements---which may be miraculous---but in this
case warns that the program may not be live.

Termination is proved by mapping the situations in a strongly connected
component to a well-founded set. Each component must be associated
with a function from the program state to $\mathtt{nat}$. Socos generates
a verification condition that the value of the termination function
strictly decreases by the loop transition. For recursive procedures,
the termination function is over the parameter list, and must be shown
to decrease by each recursive call.

\paragraph{Proof checking.}

Parallel to each generated lemma, the program checker generates a
proof script that can be executed by PVS to produce a transcript of
the proof run. Socos implements a light-weight interface to the PVS
Lisp process, through which the generated proof script is executed
and all open (unproved) sequents are collected from the proof transcript.
Socos extracts the open sequents on-line as the proof progresses,
allowing incremental extension of the proof status report. By applying
the primitive inference rules of PVS, the proof script expands the
generated correctness lemma into a proof tree where each leaf is of
the form 
\[
\gamma_{1},\dots,\gamma_{n}\vdash\delta
\]
where $\gamma_{1},\dots,\gamma_{n}$ are the assumptions from the
source situation and transition, and $\delta$ is a single constraint
from the target situation. The default proof strategy applied to each
such leaf is user-definable. The following PVS strategy, which we
will use in the case study, expands all relevant definitions in the
sequent, loads the lemmas supplied as parameters into the antecedent,
and invokes Yices as an end-game prover:

\noindent \texttt{}
\begin{lstlisting}[basicstyle={\small\ttfamily},columns=fullflexible,frame=lines,tabsize=3]
(defstep endgame (&optional (lemmas nil))
 		(let ((introduce-lemmas `(then ,@(loop for l in lemmas append `((lemma ,l))))))
 			 (then
 				(skosimp*)
 				(auto-rewrite-defs :always? t) 
 				(assert)
 				introduce-lemmas 
 				(yices) 
 				(fail)))
 		"End-game strategy" "Invoking Yices, supplying lemmas: ~{~a~^, ~}")
\end{lstlisting}

\noindent Yices either proves the lemma, or the entire strategy fails.
Definitions not expanded in the second step appear as uninterpreted
constants and the supplied lemmas as axioms to Yices. This allows
feeding specific lemmas in cases where automatic reasoning with the
definitions is infeasible; the example in Section
\ref{sec:Background-theories} demonstrates this mechanism.

\subsection{Background theories}

Socos contexts can directly import PVS background theories containing
specifications, definitions and lemmas useful for specifying and verifying
invariant diagrams. Good background theories are challenging to develop.
For a new domain we spend about half the time developing the background
theories, while the other half is spent building and verifying the
diagrams. However, the time vested in developing background theories
is typically amortized over several programs in the same domain. Background
theories can build on existing theories, for instance from the PVS
prelude or the comprehensive NASA Langley theory collection \cite{NASALIBS}.
Socos provides a small library of background theories and strategies.
It currently consists of just a few basic theories for arrays and
vectors, but we plan on extending it based on case studies.

\section{Background theories for sorting\label{sec:Background-theories}}

This section describes two background theories: $\texttt{vector}$,
introducing a type for dynamic arrays, and $\texttt{sorting}$, introducing
a set of predicates for specifying sortedness and permutations. We
will use these theories in program developed in the remainder of the
paper.

\subsection{Dynamic arrays}

PVS\emph{ dependently typed records} provide a convenient way of modeling
dynamic (resizable) arrays containing elements of the generic type
$\mathtt{T}$:
\begin{quote}
$\begin{array}[t]{@{}l}
\mathtt{vector[T:type+]}:\;\textsf{\textbf{theory}}\\
\textsf{\textbf{begin}}\\
\quad\begin{array}[t]{l}
\mathtt{vector:\textsf{\textbf{type+}}=[\#\, len\colon nat,\, elem\colon[below(len)\rightarrow T]\,\#]}\\
\mathtt{index(a:vector)}:\textsf{\textbf{type}}=\mathtt{below[len(a)]}
\end{array}
\end{array}$
\end{quote}
\noindent The $\mathtt{vector}$ type is a record type with a field
$\mathtt{len}$ for the number of elements and field $\mathtt{elem}$
for accessing the contents. The value of the field $\mathtt{elem}$
is a function whose domain depends on value of the field $\mathtt{len}$.
The type $\mathtt{below}$ is a dependent type itself, defined as
$\mathtt{below(i:nat):\textsf{\textbf{type}}=\{s:nat|s<i\}}$ in the
PVS prelude. Since PVS is a logic of total functions, $\mathtt{elem(a)}$
may only be applied within its domain; accessing $\mathtt{elem(a)}$
outside its domain will generate unprovable TCCs. The second line
introduces the shorthand $\mathtt{index(a)}$ for the domain of $\mathtt{elem(a)}$.
Access and update of an element can now be defined as:
\begin{quote}
$\begin{array}[t]{@{}l}
\quad\begin{array}[t]{l}
\mathtt{access(a\colon\mathtt{vector},i\colon index(a)):T=elem(a)(i)}\\
\mathtt{update(a\colon\mathtt{vector},i\colon index(a),x:T):vector=}\\
\quad\mathtt{(\#\, len\colon=len(a),elem\colon=elem(a)\,\textsf{\textbf{with}}\,[i\colon=x]\,\#)}
\end{array}\end{array}$
\end{quote}
In the sequel, we will write $\mathtt{a[i]}$ instead of $\mathtt{access(a,i)}$
for brevity. Finally, a predicate that two arrays are element-wise
equal on a common subrange will become useful later:
\begin{quote}
$\begin{array}[t]{@{}l}
\quad\begin{array}[t]{l}
\mathtt{eql(a\colon\mathtt{vector},b\colon\mathtt{vector},l\colon nat,r\colon nat):bool}=\\
\quad\mathtt{\forall(i:nat):l\le i\land i<r\land i<len(a)\land i<len(b)\Rightarrow a[i]=b[i]}\\
\end{array}\\
\textsf{\textbf{end}}\;\mathtt{vector}\\
\\
\end{array}$
\end{quote}

\subsection{Sortedness, permutation and swap}

We focus in the sequel on sorting arrays of type $\mathtt{vector[int]}$.
The postcondition of a sorting program should state that the array
(1) is in non-decreasing order, and (2) has preserved all values of
the original array. We introduce a predicate $\mathtt{sorted}$ to
express property (1) in a new PVS theory:
\begin{quote}
$\begin{array}[t]{@{}l}
\mathtt{sorting}:\;\textsf{\textbf{theory}}\\
\textsf{\textbf{begin}}\\
\quad\begin{array}[t]{@{}l}
\textsf{\textbf{importing}}\ \mathtt{vector[int]}\\
\mathtt{a,b,c}:\;\textsf{\textbf{var}}\ \mathtt{vector}\\
\mathtt{sorted(a):bool=\forall(i,j:index(a)):i<j\Rightarrow a[i]\le a[j]}
\end{array}
\end{array}$
\end{quote}
\noindent In the sequel we use $\mathtt{sorting}$ as a background
theory for our sorting program, extending it with additional definitions
as needed. To formalize property (2), we introduce a binary predicate
$\mathtt{perm}$, asserting the existence of a bijection over the
indexes that makes vectors $\mathtt{a}$ and $\mathtt{b}$ elementwise
equal:
\begin{quote}
$\quad\mathtt{perm(a,b):bool=\begin{array}[t]{@{}l}
\mathtt{\mathtt{\exists(f:(bijective?(index(a),index(b)))):}}\\
\quad\mathtt{\forall(i:index(b)):b[i]=a[f(i)]}
\end{array}}$
\end{quote}
\noindent For an automatic prover reasoning in terms of this definition
is problematic, since it requires demonstration of a bijection. Quantifiers
render Yices incomplete, and the catch-all strategy $\texttt{grind}$
fails to prove even that $\mathtt{perm}$ is reflexive. When verifying
algorithms which manipulate pairs of elements it is more fruitful
to consider permutation as the smallest equivalence relation that
is invariant under the pairwise swap. Proceeding in this direction,
we introduce and prove the following properties of $\mathtt{perm}$
in PVS:
\begin{quote}
\noindent $\quad\begin{array}[t]{@{}l}
\mathtt{perm\_len:\;\textsf{\textbf{lemma}}\ perm(a,b)\Rightarrow len(a)=len(b)}\\
\mathtt{perm\_ref:\;\textsf{\textbf{lemma}}\ perm(a,a)}\\
\mathtt{perm\_sym:\;\textsf{\textbf{lemma}}\ perm(a,b)\Rightarrow perm(b,a)}\\
\mathtt{perm\_trs:\;\textsf{\textbf{lemma}}\ \mathtt{perm(a,b)\land perm(b,c)\Rightarrow perm(a,c)}}
\end{array}$
\end{quote}
\noindent The first lemma states that permutations have equal length,
allowing the prover to infer that a valid index in an array is also
a valid index in any permutation of the array. The remaining lemmas
state that permutation is an equivalence relation. Proving these four
lemmas is a straightforward exercise in PVS, involving in each case
finding the right instantiation of the bijection $\mathtt{f}$. Next,
we introduce a function $\mathtt{swap}$ for exchanging the elements
at indexes $\mathtt{i}$ and $\mathtt{j}$, while keeping the remainder
of the elements in the array unchanged:
\begin{quote}
$\quad\begin{array}[t]{l}
\mathtt{swap(a,(i,j\colon index(a))):\{b|len(b)=len(a)\}=a[i\leftarrow a[j]][j\leftarrow a[i]]}\end{array}$
\end{quote}
\noindent That $\mathtt{swap}$ maintains the length is encoded in
a predicate subtype. All array manipulations in the heapsort program
will be pairwise swaps, so the $\texttt{endgame}$ strategy only needs
to know the following about $\mathtt{swap}$: the effect on subsequent
accesses, and that $\mathtt{perm}$ is maintained. We state these
properties as follows:
\begin{quote}
\noindent $\quad\begin{array}[t]{l}
\mathtt{swap\_acc:\;\textsf{\textbf{lemma}}}\\
\begin{array}[t]{l}
\quad\mathtt{\forall(a,(i,j,k:index(a))):swap(a,i,j)[k]=a[}\begin{array}[t]{l}
\mathtt{\textsf{\textbf{if}}\ k=i\ \textsf{\textbf{then}}\ j}\\
\mathtt{\textsf{\textbf{elsif}}\ k=j\ \textsf{\textbf{then}}\ i}\\
\mathtt{\textsf{\textbf{else}}\ k\ \textsf{\textbf{endif}}\ ]}
\end{array}\end{array}\\
\mathtt{swap\_perm:\;\textsf{\textbf{lemma}}}\\
\mathtt{\quad\forall(a,(i,j:index(a))):perm(a,swap(a,i,j))}
\end{array}$
\end{quote}
\noindent The proofs are trivial: the first follows directly
from the definitions, and the second by supplying the suitable bijection.
To support automatic reasoning in terms of the above more
abstract properties of $\mathtt{perm}$ and $\mathtt{swap}$ rather
than the definitions, we turn off auto-rewrites:
\begin{quote}
$\begin{array}[t]{l}
\quad\textsf{\textbf{auto\_rewrite-}}\ \mathtt{perm,\ swap}\\
\textsf{\textbf{end}}\ \mathtt{sorting}
\end{array}$
\end{quote}
\noindent This directive prevents $\mathtt{perm}$ and $\mathtt{swap}$
from being expanded, and hence they will be treated as uninterpreted
functions by Yices when $\texttt{(endgame)}$ is invoked. We ask Socos
to import the background theory and invoke the lemmas automatically
by adding the following lines to the verification context:
\begin{quote}
\noindent $\begin{array}[t]{l}
\textsf{\textbf{importing}}\ \mathtt{sorting}\\
\textsf{\textbf{strategy}}\begin{array}[t]{l}
\texttt{"(endgame :lemmas}\begin{array}[t]{l}
\texttt{(}\texttt{perm\_len perm\_ref perm\_sym}\\
\texttt{ perm\_trs swap\_acc swap\_perm))"}
\end{array}\end{array}
\end{array}$
\end{quote}

\section{Case study: heapsort\label{sec:Example:Heapsort}}

Heapsort is an in-place, comparison-based
sorting algorithm from the class of selection sorts. It achieves $O(n\,\text{log}\, n)$
worst and average case performance by storing the unsorted elements
in a binary max-heap structure, allowing for constant time retrieval
of the maximal element and logarithmic time recovery of the heap property
after the maximal element has been removed. The algorithm shown here
is the one given by Cormen et al. in \cite[Ch. 6]{580470}. It comprises
two loops in sequence. The first loop builds a max-heap out of an
unordered array by extending a partial heap one element at a time,
starting from the end of the array. The second loop maintains a sorted
subarray after the heap, and in each iteration extends the sorted
portion by swapping the root of the max-heap with the last element
of the heap, and then restores the heap property for the next iteration.

\subsection{Situation structure}

We introduce a procedure $\mathtt{heapsort}$, which given the mutable
(value-result) parameter $\mathtt{a}$ of type $\mathtt{vector[int]}$,
should achieve the postcondition $\mathtt{sorted(a)\land perm(a,a_{0})}$,
where $\mathtt{a_{0}}$ denotes the original value of $\mathtt{a}$.
We design $\mathtt{heapsort}$ around the two loops $\textsc{BuildHeap}$
and $\textsc{TearHeap}$. The former builds the heap out of the unordered
array $\mathtt{a}$ by moving in each iteration one element of the
non-heap portion of $\mathtt{a}$ into its correct place in the heap
portion; the latter then sorts $\mathtt{a}$ by selecting in each
iteration the first (root) element from the heap portion and prepending
it to the sorted portion of the array. $\textsc{TearHeap}$ is not
entered until $\textsc{BuildHeap}$ has completed, so the same loop
counter $\mathtt{k}$ can be used in both loops. In both situations
$\mathtt{k}$ will be in the range $[0..\mathtt{len(a)}]$, and $\mathtt{perm(a,a_{0})}$
is also also an invariant of both loops. 

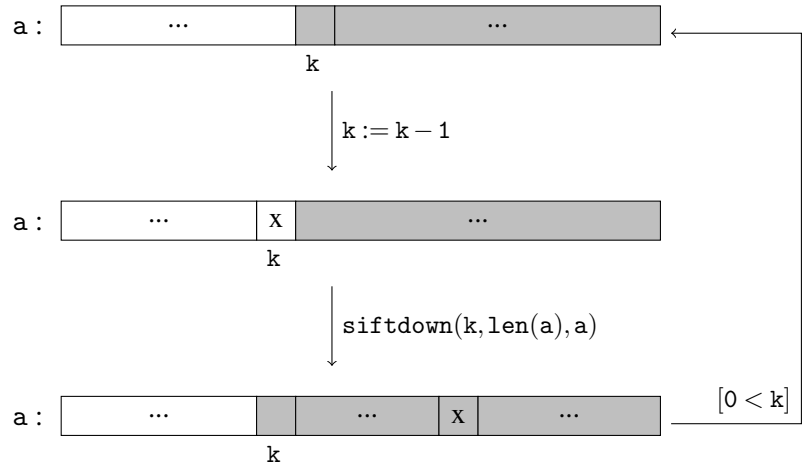
\begin{figure}
\noindent \begin{centering}
\begin{tikzpicture}
\node[arr] (i) at (0,-10ex)
{
  \node{}; \\
  \node[anchor=south] {$\mathtt{a}$ : \mbox{}}; &
  \node[rng,minimum width=18ex,minimum height=3ex]  {...}; &
  \node[elm,heap,minimum width=3ex,minimum height=3ex]  { }; &
  \node[rng,heap,minimum width=25ex,minimum height=3ex]  {...}; \\
  && \node[idx] {$\mathtt k$}; \\
};
\node[arr] (ii) at (0,-25ex)
{
  \node{}; \\
  \node[anchor=south] {$\mathtt{a}$ : \mbox{}}; &
  \node[rng,minimum width=15ex,minimum height=3ex]  {...}; &
  \node[elm,minimum width=3ex,minimum height=3ex]  {x}; &
  \node[rng,heap,minimum width=28ex,minimum height=3ex]  {...}; \\
  && \node[idx] {$\mathtt k$}; \\
};

\node[arr] (iii) at (0,-40ex)
{
  \node{}; \\
  \node[anchor=south] {$\mathtt{a}$ : \mbox{}}; &
  \node[rng,minimum width=15ex,minimum height=3ex]  {...}; &
  \node[elm,heap,minimum width=3ex,minimum height=3ex]  { }; &
  \node[rng,heap,minimum width=11ex,minimum height=3ex]  {...}; &
  \node[elm,heap,minimum width=3ex,minimum height=3ex]  {x}; &
  \node[rng,heap,minimum width=14ex,minimum height=3ex]  {...}; \\
  && \node[idx] {$\mathtt k$}; \\
};

\draw[->] (i.south) -- node[right,stmt] {$\mathtt{k:=k-1}$} (i.south|-ii.north);

\draw[->] (ii.south) -- node[right,stmt] {$\mathtt{siftdown(k,len(a),a)}$} (ii.south|-iii.north);

\draw[->] (iii.east) -- +(10ex,0ex) node[stmt,above left] {$\mathtt{[0<k]}$} |- (i.east);

\end{tikzpicture}
\par\end{centering}

\caption[Building the heap]{Building the heap. The shaded portion satisfies the max-heap property\label{fig:Building-the-heap}}
\end{figure}

In $\textsc{BuildHeap}$, the heap is extended leftwards one element
at a time by decreasing $\mathtt{k}$. The portion to the right of
$\mathtt{k}$ satisfies the following \emph{max-heap property}: an
element at index $i$ is greater than or equal to both the element
at index $2i+1$ (the {}``left child'') and the element at index
$2i+2$ (the {}``right child''). Figure \ref{fig:Building-the-heap}
shows the invariant of $\textsc{BuildHeap}$ and the loop transition.
The loop terminates when $\mathtt{k}$ reaches zero. For each iteration,
after $\mathtt{k}$ has been decremented the new element at position
$\mathtt{k}$ must be {}``sifted down'' into the heap to re-establish
the max-heap property. We defer this task to another procedure, $\mathtt{siftdown}$,
which is to be implemented in the next section. The parameters to
$\mathtt{siftdown}$ are the left and right bounds of the heap, as
well as the array itself. 

We now formalize the heap property. We extend the $\mathtt{sorting}$
background theory with functions $\mathtt{l}$ and $\mathtt{r}$ for
the index of the left and right child respectively, and a predicate
$\mathtt{heap}$ expressing that a subrange of $\mathtt{a}$ satisfies
the max-heap property:
\begin{quote}
$\begin{array}[t]{l}
\mathtt{l(i\colon nat):nat=2\times i+1}\\
\mathtt{r(i\colon nat):nat=2\times i+2}\\
\mathtt{heap(a,(m,n:nat)):bool=m\le n\land n\le len(a)\ \land}\begin{array}[t]{l}
(\mathtt{\forall(i:nat):m\le i\Rightarrow}\\
\quad\mathtt{(l(i)<n\Rightarrow a[i]\ge a[l(i)])\land}\\
\quad\mathtt{(r(i)<n\Rightarrow a[i]\ge a[r(i)])})
\end{array}
\end{array}$
\end{quote}
\noindent We get that $\textsc{BuildHeap}$ should maintain $\mathtt{heap(a,k,len(a))}$.
When the loop terminates, $\mathtt{heap(a,0,len(a))}$ should hold.

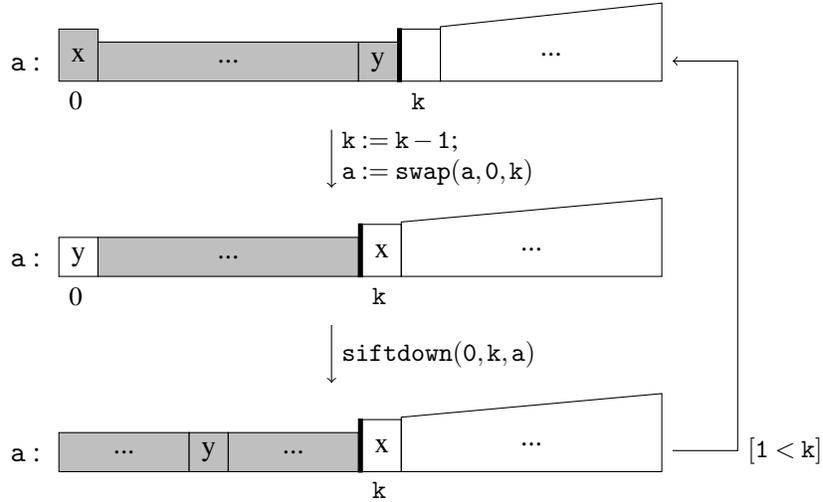
\begin{figure}
\noindent \begin{centering}
\begin{tikzpicture}
\node[arr] (i) at (0,-15ex)
{
  \node[anchor=south] {$\mathtt{a}$ : \mbox{}}; &
  \node[elm,heap,minimum width=3ex,minimum height=4ex] {x}; &
  \node[rng,heap,minimum width=20ex,minimum height=3ex]  {...}; &
  \node[elm,heap,minimum width=3ex,minimum height=3ex]  {y}; &
  \node[sep,minimum height=4ex] {}; &
  \node[elm,minimum width=3ex,minimum height=4ex]  {}; &
 \node[rng,trapezium,trapezium stretches body,trapezium left angle=0,trapezium right angle=60,minimum width=6ex,minimum height=17ex,shape border rotate=90] {...}; \\
  & \node[idx] {$0$}; & & & & \node[idx] {$\mathtt k$}; \\
};
\node[arr] (ii) at (0,-30ex)
{
  \node[anchor=south] {$\mathtt{a}$ : \mbox{}}; &
  \node[elm,minimum width=3ex,minimum height=3ex] {y}; &
  \node[rng,heap,minimum width=20ex,minimum height=3ex] {...}; &
  \node[sep,minimum height=4ex] {}; &
  \node[elm,minimum width=3ex,minimum height=4ex] {x}; &
  \node[rng,trapezium,trapezium stretches body,trapezium left angle=0,trapezium right angle=60,minimum width=6ex,minimum height=20ex,shape border rotate=90] {...}; \\
  & \node[idx] {$0$}; & & & \node[idx] {$\mathtt k$}; \\
};
\node[arr] (iii) at (0,-45ex)
{
  \node[anchor=south] {$\mathtt{a}$ : \mbox{}}; &
  \node[rng,heap,minimum width=10ex,minimum height=3ex]  {...}; &
  \node[elm,heap,minimum width=3ex,minimum height=3ex]  {y}; &
  \node[rng,heap,minimum width=10ex,minimum height=3ex]  {...}; &
  \node[sep,minimum height=4ex] {}; &
  \node[elm,minimum width=3ex,minimum height=4ex] {x}; &
  \node[rng,trapezium,trapezium stretches body,trapezium left angle=0,trapezium right angle=60,minimum width=6ex,minimum height=20ex,shape border rotate=90] {...}; \\
  & & & & & \node[idx] {$\mathtt k$}; \\
};


\draw[->] (i.south) -- node[right,stmt] {\begin{minipage}{10ex}$\mathtt{k:=k-1;}$\\$\mathtt{a:=swap(a,0,k)}$\end{minipage}} (i.south|-ii.north);

\draw[->] (ii.south) -- node[right,stmt] {$\mathtt{siftdown(0,k,a)}$} (ii.south|-iii.north);

\draw[->] (iii.east) -- +(5ex,0ex) node[stmt,right] {$\mathtt{[1<k]}$} |- (i.east);


\end{tikzpicture}
\par\end{centering}

\caption[Sorting the array]{Sorting the array. The shaded portion satisfies the max-heap property,
the sloping portion is sorted, and the array is partitioned around
$\mathtt{k}$\label{fig:Tearing-the-heap}}
\end{figure}

In situation $\textsc{TearHeap}$, which is entered after $\textsc{BuildHeap}$
has completed, we again iterate leftwards, now maintaining the heap
to the left of $\mathtt{k}$, and a sorted subarray to the right of
$\mathtt{k}$. The loop is iterated while $\mathtt{k}>1$ (when the
heap contains a single element, the array is already sorted). In each
iteration, $\mathtt{k}$ is decremented, then the element at index
$\mathtt{k}$ element is exchanged with the element at index $0$
(the root of the heap) to extend the sorted portion. As the leftmost
portion may no longer be a heap, this is followed by a call to $\mathtt{siftdown}$
to restore the heap property. Additionally, to infer that the extended
right portion is sorted, we also need to know that the array is \emph{partitioned}
around $\mathtt{k}$, i.e., that the elements to the left of $\mathtt{k}$
are smaller than or equal to the elements to the right of (and at)
$\mathtt{k}$. An informal diagram for the $\textsc{TearHeap}$ situation
and the intermediate states in the loop transition is shown in Figure
\ref{fig:Tearing-the-heap}. In this figure we have indicated with
sloping that a portion of the array is sorted in non-decreasing order.

\begin{figure}
\noindent \begin{centering}
\begin{tikzpicture}
  \node[procedure] (heapsort) at (0ex,0ex) {
    \begin{minipage}{70ex}
      \procedurelabel{heapsort} [ \ibpkw{valres} \declaration{\ibpid{a} \colon \ibpid{vector[int]}} ]\\
      \procedurerule\\
      \invariant{k \colon \ibpkw{pvar}\;nat} \\
     \procedurerule\\
     \vspace{28ex}
   \end{minipage}
  };
  \node[postcondition] (Post) at (54ex,-9ex) {
      \begin{minipage}{12ex}
        \invariant{sorted(a)} \\
        \invariant{perm(a,a_0)}
      \end{minipage}
    };
  \node[situation] (Constraints) at (1ex,-9ex) {
      \begin{minipage}{50ex}
        \situationlabel{Constraints} \\
        \situationrule \\
        \invariant{perm(a,a_0)} \\
        \invariant{k \le len(a)} \\
	    \vspace{15ex}
      \end{minipage}
    };
  \node[situation] (BuildHeap) at (2ex,-20ex) {
      \begin{minipage}{20ex}
        \situationlabel{BuildHeap} \\
        \situationrule \\
        \invariant{heap(a,k,len(a))} 
      \end{minipage}
    };
  \node[situation] (TearHeap) at (28ex,-20ex) {
      \begin{minipage}{20ex}
        \situationlabel{TearHeap} \\
        \situationrule \\
        \invariant{partitioned(a,k)} \\
        \invariant{sorted(a,k)} \\
        \invariant{heap(a,0,k)} 
    \end{minipage}
    };

\end{tikzpicture}
\par\end{centering}

\caption{Heapsort situations\label{fig:Heapsort-skeleton}}
\end{figure}
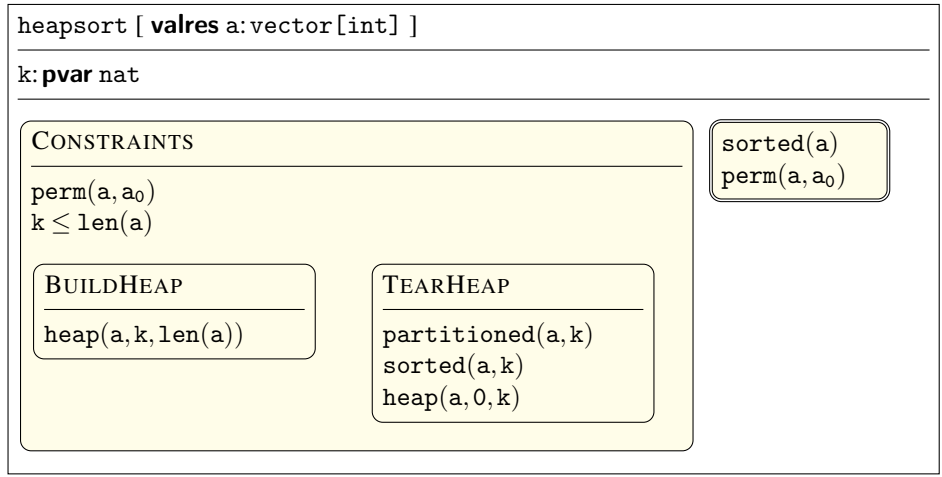

To be able to express the constraints of $\textsc{TearHeap}$ concisely
we introduce two predicates into the background theory; one expressing
that the rightmost portion of an array is sorted, and one that an
array is partitioned around a given index:
\begin{quote}
$\begin{array}[t]{l}
\mathtt{sorted(a,(n \colon  upto(len(a)))):bool=\forall(i,j \colon index(a)):n\le i\land i<j\Rightarrow a[i]\le a[j]}\\
\mathtt{partitioned(a,(k \colon upto(len(a)))):bool=\forall(i,j \colon index(a)):i<k\land k\le j\Rightarrow a[i]\le a[j]}
\end{array}$
\end{quote}
\noindent With these declarations added to the background theory,
we can now give a first situation structure for the procedure $\mathtt{heapsort}$.
A partial invariant diagram is shown in Figure \ref{fig:Heapsort-skeleton}.
Since $\textsc{Constraints}$ is also over the local variable $\mathtt{k}$,
the postcondition cannot be nested inside $\textsc{Constraints}$;
hence we have repeated the constraint $\mathtt{perm(a,a_{0})}$ in
the postcondition.

\subsection{Loop initialization and exit}

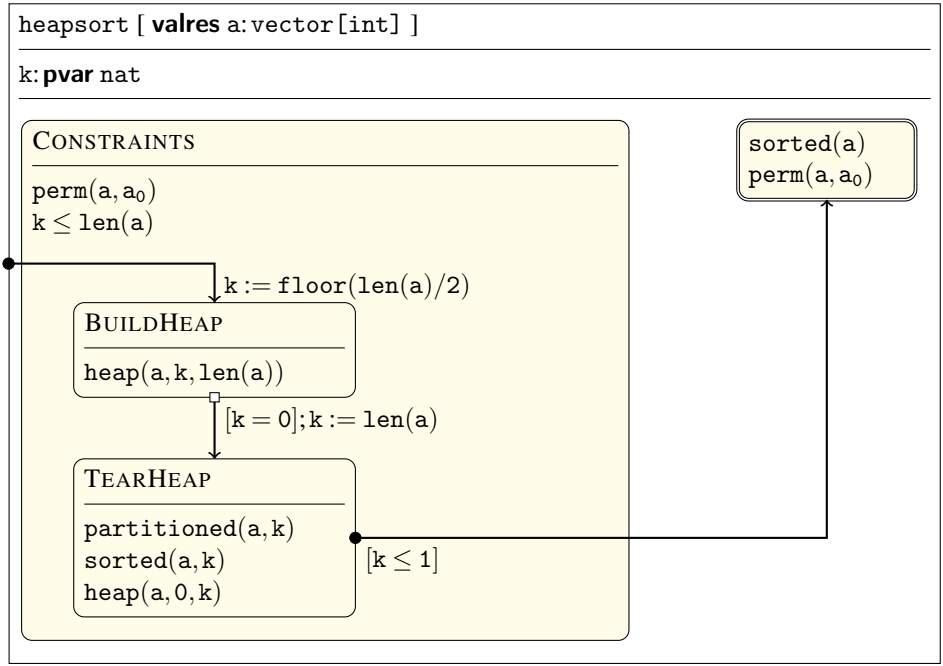
\begin{figure}
\noindent \begin{centering}
\begin{tikzpicture}
  \node[procedure] (heapsort) at (0ex,0ex) {
    \begin{minipage}{70ex}
      \procedurelabel{heapsort} [ \ibpkw{valres} \declaration{\ibpid{a} \colon \ibpid{vector[int]}} ]\\
      \procedurerule\\
      \invariant{k \colon \ibpkw{pvar}\;nat} \\
     \procedurerule\\
     \vspace{44ex}
   \end{minipage}
  };
  \node[postcondition] (Post) at (56ex,-9ex) {
      \begin{minipage}{12ex}
        \invariant{sorted(a)} \\
        \invariant{perm(a,a_0)}
      \end{minipage}
    };
  \node[situation] (Constraints) at (1ex,-9ex) {
      \begin{minipage}{45ex}
        \situationlabel{Constraints} \\
        \situationrule \\
        \invariant{perm(a,a_0)} \\
        \invariant{k \le len(a)} \\
	    \vspace{31ex}
      \end{minipage}
    };
  \node[situation] (BuildHeap) at (5ex,-23ex) {
      \begin{minipage}{20ex}
        \situationlabel{BuildHeap} \\
        \situationrule \\
        \invariant{heap(a,k,len(a))} 
      \end{minipage}
    };
  \node[situation] (TearHeap) at (5ex,-35ex) {
      \begin{minipage}{20ex}
        \situationlabel{TearHeap} \\
        \situationrule \\
        \invariant{partitioned(a,k)} \\
        \invariant{sorted(a,k)} \\
        \invariant{heap(a,0,k)} 
    \end{minipage}
    };

  \draw[transition,->] (heapsort.north west) ++(0ex,-20ex) node[blackdot] {} -| node[statement] {$\mathtt{k:=floor(len(a)/2)}$} (BuildHeap.north) ;
 
  \draw[transition,->] (BuildHeap.south) node[choice] {} node[statement] {$\mathtt{[k=0];k:=len(a)}$} -- (TearHeap.north) ;
  
  \draw[transition,->] (TearHeap.east) node[blackdot] {} node[statement] {$\mathtt{[k \le 1]}$} -| (Post.south) ;
\end{tikzpicture}
\par\end{centering}

\caption{$\texttt{heapsort}$ with acyclic transitions in place\label{fig:Heapsort-initial-final-trans}}
\end{figure}

Since the initial and final transitions, as well as the transition
between $\textsc{BuildHeap}$ and $\textsc{TearHeap}$ do not depend on
the $\texttt{siftdown}$ procedure, they can be added and checked
immediately. We first consider the initial transition. While we could
initialize the loop counter to $\mathtt{len(a)}$, we can do better.
$\mathtt{heap(a,m,len(a))}$ is actually true for any index
$\mathtt{m}$ on the bottom level of the heap, i.e., satisfying
$\mathtt{\lfloor len(a)/2\rfloor\le m}$.  We can confirm this
hypothesis by adding the statement $\mathtt{k:=floor(len(a)/2)}$ as
the initial transition and asking Socos to check
$\mathtt{heapsort}$. Socos responds that all transitions are
consistent, and also points out that the procedure is not live. We
proceed by adding the two exit transitions: from $\textsc{BuildHeap}$
to $\textsc{TearHeap}$, and from $\textsc{TearHeap}$ to the
postcondition.  The updated diagram is shown in Figure
\ref{fig:Heapsort-initial-final-trans}.  Rechecking the program, Socos
confirms that the program is consistent (but still not live). However,
before we can add the loop transitions, we need to implement and
verify $\texttt{siftdown}$.

\subsection{The $\texttt{siftdown}$ procedure}

\begin{figure}
\noindent \begin{centering}
\begin{tikzpicture}
  \node[procedure] (siftdown) at (0ex,0ex) {
    \begin{minipage}{55ex}
      \procedurelabel{siftdown} [ \declaration{\ibpid{m} \colon \ibpid{nat}, \; \ibpid{n} \colon \ibpid{nat} \ \  \ibpkw{valres} \; \ibpid{a} \colon \ibpid{vector[int]}}  ]\\
      \procedurerule\\
     \procedurerule\\
     \vspace{15ex}
   \end{minipage}
  };

  \node[precondition] (Pre) at (2ex,-7ex) {
      \begin{minipage}{18ex}
        \invariant{m \le n \land n \le len(a)} \\
        \invariant{heap(a,m+1,n)}
      \end{minipage}
   };

  \node[postcondition] (Post) at (30ex,-7ex) {
      \begin{minipage}{20ex}
        \invariant{heap(a,m,n)} \\
        \invariant{perm(a,a_0)} \\
        \invariant{eql(a,a_0,0,m)} \\
        \invariant{eql(a,a_0,n,len(a))} \\
      \end{minipage}
   };

\end{tikzpicture}
\par\end{centering}

\caption{$\texttt{siftdown}$ specification\label{fig:Siftdown-spec}}
\end{figure}
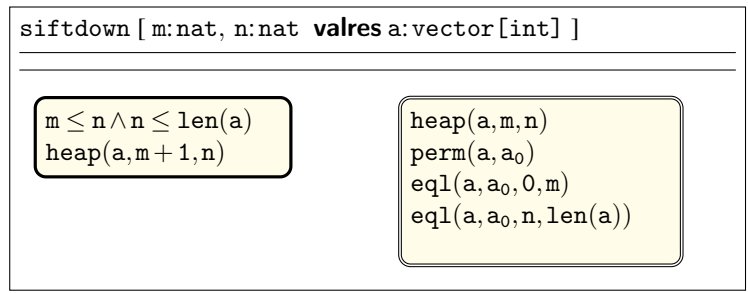

The parameters to $\texttt{siftdown}$ are the left bound $\mathtt{m}$,
the right bound $\mathtt{n}$, and the array $\mathtt{a}$. Assuming
the subrange $[\mathtt{m}+1..\mathtt{n})$ satisfies the heap property,
$\texttt{siftdown}$ should ensure upon completion that the subrange
$[\mathtt{m}..\mathtt{n})$ satisfies the heap property, that the
subranges $[\mathtt{0}..\mathtt{m})$ and $[\mathtt{n}..\mathtt{len(a)})$
are unchanged, and that the updated array is a permutation of the
original array. A pre-post specification is given in Figure \ref{fig:Siftdown-spec}.

\begin{figure}[h]
\noindent \begin{centering}
\begin{tikzpicture} 
[hp/.style={heap,draw,shape=circle,minimum size=3.25ex,inner sep=0pt,font=\small},
 nd/.style={draw,thin,shape=circle,minimum size=3.25ex,inner sep=0pt,font=\small},
 level 2/.style={sibling distance=12ex},
 level 3/.style={sibling distance=6ex}
]

\node[hp] (p1) at (22ex,0ex) {} [level distance=4ex] 
 child { node[nd] (x1) {} 
         child { node[hp] (a1) {A} 
                 child { node[hp] (c1) {C} }
                 child { node[hp] (d1) {D} } }
         child { node[hp] (b1) {B} 
                 child { node[hp] (e1) {E} }
                 child { node[hp] (f1) {F} } } };
\node[fit=(p1)(x1)(a1)(b1)(c1)(d1)(e1)(f1)] (s1) {};
\draw (x1.south) node[anchor=north] {\small $\mathtt{k}$};
\draw (a1.west) node[anchor=east] {\small $\mathtt{l(k)}$};
\draw (b1.east) node[anchor=west] {\small $\mathtt{r(k)}$};
\draw[dashed] (p1.west) to [bend right=30] node[above,sloped]{$\le$} (a1.north) ;
\draw[dashed] (p1.east) to [bend left=30] node[above,sloped]{$\ge$} (b1.north) ;

\node[hp] (p2) at (0ex,-30ex) {} [level distance=4ex] 
 child { node[hp] (a2) {A} 
         child { node[nd] (x2) {} 
                 child { node[hp] (c2) {C} }
                 child { node[hp] (d2) {D} } }
         child { node[hp] (b2) {B} 
                 child { node[hp] (e2) {E} }
                 child { node[hp] (f2) {F} } } };
\node[fit=(p2)(x2)(a2)(b2)(c2)(d2)(e2)(f2)] (s2) {};
\draw (x2.south) node[anchor=north] {\small $\mathtt{k}$};
\draw[dashed] (a2.west) to [bend right=45] node[above,sloped]{$\le$} (c2.north) ;
\draw[dashed] (a2.south) to [bend left=25] node[above,sloped]{$\le$} (d2.north) ;

\node[hp] (p3) at (44ex,-30ex) {} [level distance=4ex] 
 child { node[hp] (b3) {B} 
         child { node[hp] (a3) {A} 
                 child { node[hp] (c3) {C} }
                 child { node[hp] (d3) {D} } }
         child { node[nd] (x3) {} 
                 child { node[hp] (e3) {E} }
                 child { node[hp] (f3) {F} } } };
\node[fit=(p3)(x3)(a3)(b3)(c3)(d3)(e3)(f3)] (s3) {};
\draw (x3.south) node[anchor=north] {\small $\mathtt{k}$};
\draw[dashed] (b3.south) to [bend right=25] node[above,sloped]{$\ge$} (e3.north) ;
\draw[dashed] (b3.east) to [bend left=45] node[above,sloped]{$\ge$} (f3.north) ;

\draw (s1.south) -- 
  node[right,font=\small] {
    $\begin{array}{l}
       \mathtt{[r(k) < n \; \land } \\
       \mathtt{\;(a[k]<a[ l(k)] \lor a[k]<a[r(k)])]}
     \end{array}$
  } +(0,-5ex) node[inner sep=0pt] (tmp) {};

\draw[->] (tmp) --
  node[below,font=\small] {
      $\begin{array}{l}
         \mathtt{[a[r(k)] \le a[l(k)]];} \\
         \mathtt{a := swap(a,k,l(k));} \\
         \mathtt{k := l(k)} \\
      \end{array}$      
  } (tmp-|s2.north)  -- (s2.north);

\draw[->] (tmp) -- 
  node[below,font=\small] {
      $\begin{array}{l}
         \mathtt{[a[l(k)] \le a[r(k)]];} \\
         \mathtt{a := swap(a,k,r(k));} \\
         \mathtt{k := r(k)} \\
      \end{array}$      
  } (tmp-|s3.north) -- (s3.north);

\draw[->] (s2.west) -- +(-1ex,0) |- (s1.west);
\draw[->] (s3.east) -- +(1ex,0)|- (s1.east);

\end{tikzpicture}
\par\end{centering}

\caption{The $\mathtt{siftdown}$ loop invariant \label{fig:Siftdown-loop}}
\end{figure}
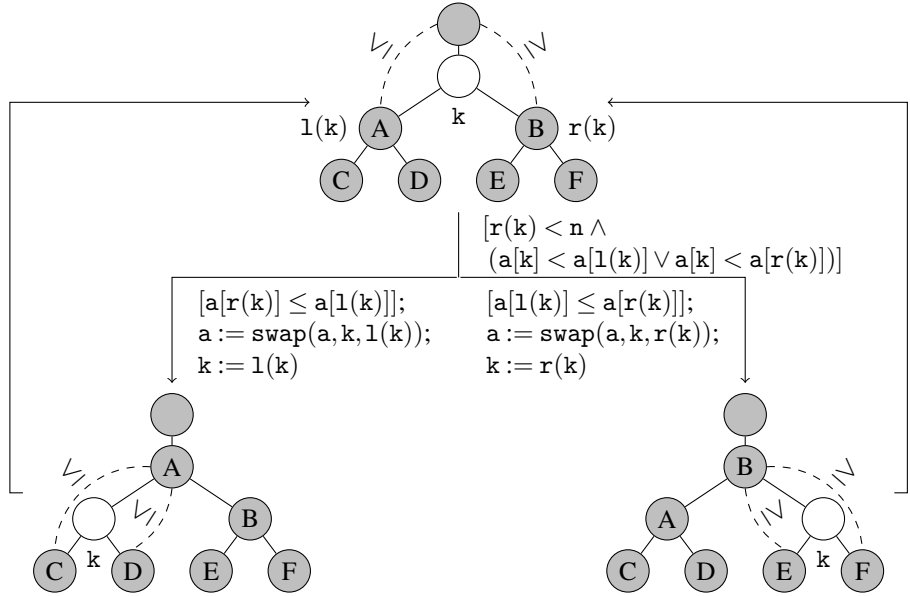

\noindent The procedure $\texttt{siftdown}$ achieves its postcondition
by {}``sifting'' the first element in the range downward into the
heap until it is either greater than or equal to both its left and
right child, or the bottom of the heap has been reached. When either
condition is true, the heap property has been restored. Each iteration
of the loop swaps the current element with the greater of its children,
maintaining the invariant that each element within the heap range,
except the current one, is greater than or equal to both its children.
The loop statement, using a counter $\mathtt{k}$ pointing to the
current element, is given in Figure \ref{fig:Siftdown-loop} together
with an illustration of the loop invariant. 
 In this figure circles represent elements within the heap range.
A shaded circle indicates that an element is known to be greater than
or equal to its children. The dashed lines indicate that the parent
of $\mathtt{k}$ is also be known to be greater than or equal to $\mathtt{k}$:s
children. This part of the invariant is required to prove that the
max-heap property holds for the new parent of $\mathtt{k}$ after
swapping. That it is maintained follows from the fact that the child
selected for swapping is known to be greater than or equal to its
children.

\begin{figure}
\noindent \begin{centering}
\begin{tikzpicture}
  \node[procedure] (siftdown) at (0ex,0ex) {
    \begin{minipage}{70ex}
      \procedurelabel{siftdown} [ \declaration{\ibpid{m} \colon \ibpid{nat}, \; \ibpid{n} \colon \ibpid{nat} \ \  \ibpkw{valres} \; \ibpid{a} \colon \ibpid{vector[int]}} ]\\
     \procedurerule\\
     \declaration{k:\ibpkw{pvar} \; nat ;} \\
     \procedurerule\\
     \vspace{53ex}
   \end{minipage}
  };

  \node[precondition] (Pre) at (2ex,-10ex) {
      \begin{minipage}{18ex}
        \invariant{m \le n \land n \le len(a)} \\
        \invariant{heap(a,m+1,n)}
      \end{minipage}
   };

  \node[postcondition] (Post) at (45ex,-10ex) {
      \begin{minipage}{20ex}
        \invariant{heap(a,m,n)} \\
        \invariant{perm(a,a_0)} \\
        \invariant{eql(a,a_0,0,m)} \\
        \invariant{eql(a,a_0,n,len(a))} \\
      \end{minipage}
   };
  
  \node[situation] (Sift) at (2ex,-25ex) {
      \begin{minipage}{35ex}
        \situationlabel{Sift} \\
        \situationrule \\
        \invariant{perm(a,a_0)} \hfill \invariant{| n-k}\\
        \invariant{m \le k \land k \le n \land n \le len(a)} \\
        \invariant{eql(a,a_0,0,m)} \\
        \invariant{eql(a,a_0,n,len(a))} \\
        \invariant{
\forall \begin{array}[t]{@{}l}
\mathtt{(i \colon nat) \colon m\le i \Rightarrow }\\
\quad\mathtt{(i \ne k\Rightarrow}\\
\quad\quad\mathtt{(l(i)<n\Rightarrow a[l(i)]\le a[i])\land}\\
\quad\quad\mathtt{(r(i)<n\Rightarrow a[r(i)]\le a[i])) \land} \\
\quad\mathtt{(l(i)=k\vee r(i)=k\Rightarrow}\\
\quad\quad\mathtt{(l(k)<n\Rightarrow a[l(k)]\le a[i])\land}\\
\quad\quad\mathtt{(r(k)<n\Rightarrow a[r(k)]\le a[i]))}\end{array}
        }
      \end{minipage}
   };

\draw[transition,->] (Pre.south) node[blackdot]{} node[statement] {$\mathtt{k:=m}$} -- (Sift.north-|Pre.south);
  
\draw[transition,->] (Sift.north) +(5ex,0) node[if]{}  |- 
node[statement,above,anchor=south west] {$
[\begin{array}[t]{l}
\mathtt{n \le r(k) \lor} \\
\mathtt{(a[l(k)]\le a[k]\land } \\
\mathtt{\;a[r(k)]\le a[k])} \; ]\\
\end{array}
$} (Post.west);

\draw[transition] (Sift.north east) ++(0,-2ex) node[if]{} -- ++(25ex,0)
node[statement,below left] {$
[\begin{array}[t]{l}
\mathtt{r(k) < n \land} \\
\mathtt{(a[k] < a[l(k)] \lor } \\
\mathtt{\;a[k] < a[r(k)])} \; ]\\
\end{array}
$}
-- ++(0,-12ex) node[if] (if1) {};

\draw[transition,->>] (if1) --
node[statement,below] {$
\begin{array}[t]{l}
\mathtt{[a[r(k)] \le a[l(k)]];} \\
\mathtt{a:=swap(a,k,l(k));} \\
\mathtt{k:=l(k)}\\
\end{array}
$}
++(-25ex,0);

\draw[transition,->>] (if1) -- ++(0,-11ex) --
node[statement,below] {$
\begin{array}[t]{l}
\mathtt{[a[l(k)] \le a[r(k)]];} \\
\mathtt{a:=swap(a,k,r(k));} \\
\mathtt{k:=r(k)}\\
\end{array}
$}
++(-25ex,0);

\end{tikzpicture}
\par\end{centering}

\caption{A first attempt at $\texttt{siftdown}$ \label{fig:Siftdown-incomplete}}
\end{figure}
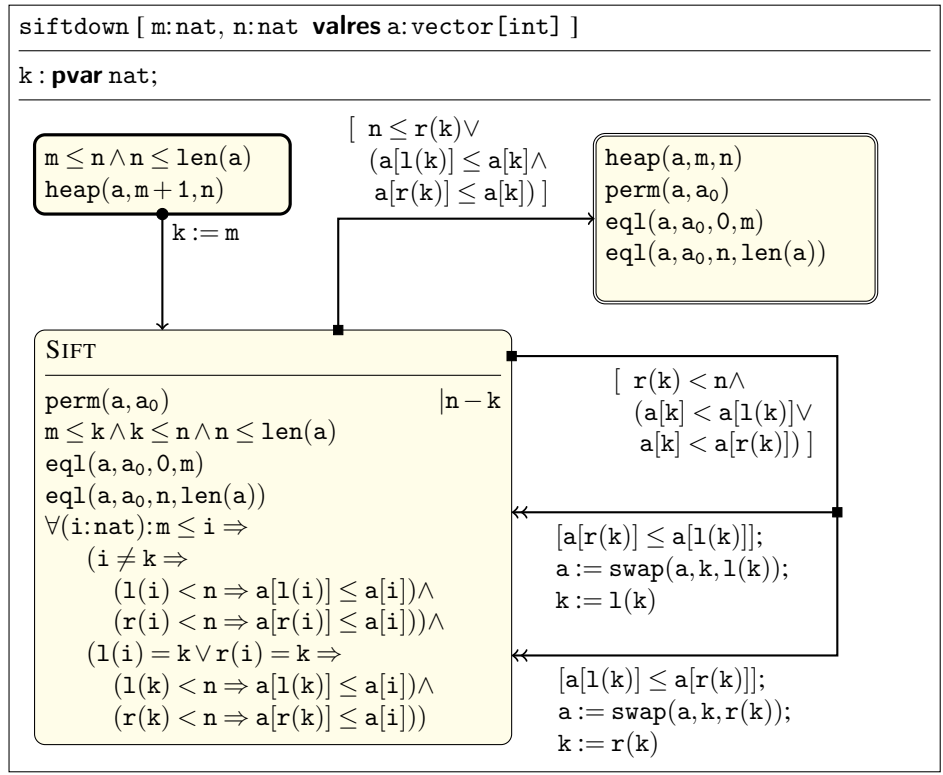

The procedure should return when either the values of both children
are less than or equal the current element, or there are no more children
within the range of the heap. More precisely, the loop should exit
to the postcondition when the following condition holds:
\begin{quote}
$\mathtt{n\le r(k)\vee(a[l(k)]\le a[k]\land a[r(k)]\le a[k])}$
\end{quote}
\noindent Figure \ref{fig:Siftdown-incomplete} shows a diagram with
an intermediate situation $\textsc{Sift}$ and the entry, loop and
exit transitions in place. The termination function $\mathtt{n-k}$
is decreased by both loop transitions.

When we check the program, Socos proves all transitions
except the exit transition; the unproved condition is shown in Figure
\ref{fig:Unproved-condition-Sift}. 
\begin{figure}
\begin{lstlisting}[basicstyle={\small\ttfamily},frame=lines]
    [-1]  n <= r(k) OR
           (a[l(k)] <= a[k] AND a[r(k)] <= a[k])
    [-2]  (n <= r(k) OR
            (a[l(k)] <= a[k] AND
              a[r(k)] <= a[k]))
           OR
           (r(k) < n AND
             (a[k] < a[l(k)] OR a[k] < a[r(k)]))
    [-3]  (perm(a, a_0))
    [-4]  m <= k and k <= n and n <= len(a)
    [-5]  eql(a, a_0, 0, m)
    [-6]  eql(a, a_0, n, len(a))
    [-7]  FORALL (i: nat):
            m <= i =>
             (i /= k =>
               (l(i) < n => a[l(i)] <= a[i]) AND
                (r(i) < n => a[r(i)] <= a[i]))
              AND
              ((l(i) = k OR r(i) = k) =>
                (l(k) < n => a[l(k)] <= a[i]) AND
                 (r(k) < n => a[r(k)] <= a[i]))
      |-------
    [1}   (heap(a, m, n))
\end{lstlisting}

\caption[Unproven condition for the exit transition from $\textsc{Sift}$.]{Unproven condition for the exit transition from $\textsc{Sift}$
\label{fig:Unproved-condition-Sift}}
\end{figure} 
The automatic strategy was unable to assert that
$\mathtt{heap(a,m,n)}$ is established by the exit transition. The
assumptions are, in fact, not strong enough to show that
$\mathtt{heap(a,m,n)}$ is maintained.  This is due to an omission of a
corner case in the program in Figure \ref{fig:Siftdown-incomplete}:
when $\mathtt{n=r(k)}$, nothing is known about the relation between
$\mathtt{a[k]}$ and $\mathtt{a[l(k)]}$.  The corner case occurs when
the left child of the current element is the last element in the heap
range, and the right child falls just outside of the heap range.  This
bug is hard to spot, and is easily missed even with extensive testing.

\begin{figure}
\noindent \begin{centering}
\begin{tikzpicture}
  \node[procedure] (siftdown) at (0ex,0ex) {
    \begin{minipage}{70ex}
      \procedurelabel{siftdown} [ \declaration{\ibpid{m} \colon \ibpid{nat}, \; \ibpid{n} \colon \ibpid{nat} \ \  \ibpkw{valres} \; \ibpid{a} \colon \ibpid{vector[int]}}  ]\\
      \procedurerule\\
     \declaration{k:\ibpkw{pvar} \; nat ;} \\
     \procedurerule\\
     \vspace{68ex}
   \end{minipage}
  };

  \node[precondition] (Pre) at (2ex,-10ex) {
      \begin{minipage}{18ex}
        \invariant{m \le n \land n \le len(a)} \\
        \invariant{heap(a,m+1,n)}
      \end{minipage}
   };

  \node[postcondition] (Post) at (25ex,-10ex) {
      \begin{minipage}{20ex}
        \invariant{heap(a,m,n)} \\
        \invariant{perm(a,a_0)} \\
        \invariant{eql(a,a_0,0,m)} \\
        \invariant{eql(a,a_0,n,len(a))} \\
      \end{minipage}
   };
  
  \node[situation] (Sift) at (2ex,-25ex) {
      \begin{minipage}{35ex}
        \situationlabel{Sift} \\
        \situationrule \\
        \invariant{perm(a,a_0)} \hfill \invariant{| n-k}\\
        \invariant{m \le k \land k \le n \land n \le len(a)} \\
        \invariant{eql(a,a_0,0,m)} \\
        \invariant{eql(a,a_0,n,len(a))} \\
        \invariant{
\forall \begin{array}[t]{@{}l}
\mathtt{(i \colon nat) \colon m\le i \Rightarrow }\\
\quad\mathtt{(i \ne k\Rightarrow}\\
\quad\quad\mathtt{(l(i)<n\Rightarrow a[l(i)]\le a[i])\land}\\
\quad\quad\mathtt{(r(i)<n\Rightarrow a[r(i)]\le a[i])) \land} \\
\quad\mathtt{(l(i)=k\vee r(i)=k\Rightarrow}\\
\quad\quad\mathtt{(l(k)<n\Rightarrow a[l(k)]\le a[i])\land}\\
\quad\quad\mathtt{(r(k)<n\Rightarrow a[r(k)]\le a[i]))}\end{array}
        }
     \vspace{10ex}
     \end{minipage}
   };

\draw[transition,->] (Pre.south) node[blackdot]{} node[statement] {$\mathtt{k:=m}$} -- (Sift.north-|Pre.south);
  
\draw[transition] (Sift.north east) ++(0,-5ex) node[if]{}  
 node[statement] {$
[\begin{array}[t]{l}
\mathtt{n \le r(k) \lor} \\
\mathtt{(a[l(k)]\le a[k]\land } \\
\mathtt{\;a[r(k)]\le a[k])} \; ]\\
\end{array}
$} -- ++(17ex,0) node[if] (if2) {} ;

\draw[transition,->] (if2)   |-
node[statement,below left] {$\mathtt{[n \ne r(k)]}$}
  (Post.south east);

\draw[transition] (if2)  -- 
node[statement,above] {$\mathtt{[n = r(k)]}$}
++(14ex,0) --  ++(0,12ex) node[if] (if3) {};

\draw[transition,->] (if3) -- ++(0,5ex) |-
node[statement,below left] 
{$
\begin{array}[t]{l}
\mathtt{[ a[k] < a[l(k)] ];} \\
\mathtt{a := swap(a,k,l(k))}
\end{array}
$} 
(Post.north east);

\draw[transition,->] (if3) --
node[statement,below] {$\mathtt{[a[l(k)] \le a[k]]}$}
 (if3-|Post.east);

\draw[transition] (Sift.north east) ++(0,-16ex) node[if]{} --
node[statement,below] {$
[\begin{array}[t]{l}
\mathtt{r(k) < n \land} \\
\mathtt{(a[k] < a[l(k)] \lor } \\
\mathtt{\;a[k] < a[r(k)])} \; ]\\
\end{array}
$}
++(25ex,0)-- ++(0,-11ex) node[if] (if1) {};

\draw[transition,->>] (if1) --
node[statement,below] {$
\begin{array}[t]{l}
\mathtt{[a[r(k)] \le a[l(k)]];} \\
\mathtt{a:=swap(a,k,l(k));} \\
\mathtt{k:=l(k)}\\
\end{array}
$}
++(-25ex,0);

\draw[transition,->>] (if1) -- ++(0,-11ex) --
node[statement,below] {$
\begin{array}[t]{l}
\mathtt{[a[l(k)] \le a[r(k)]];} \\
\mathtt{a:=swap(a,k,r(k));} \\
\mathtt{k:=r(k)}\\
\end{array}
$}
++(-25ex,0);

\end{tikzpicture}
\par\end{centering}

\caption[Final $\texttt{siftdown}$ program]{Final $\texttt{siftdown}$
  program, with corrected exit transition.  The corner case
  $\mathtt{n=r(k)}$ is handled in a separate exit
  transition\label{fig:Siftdown-final}}
\end{figure}
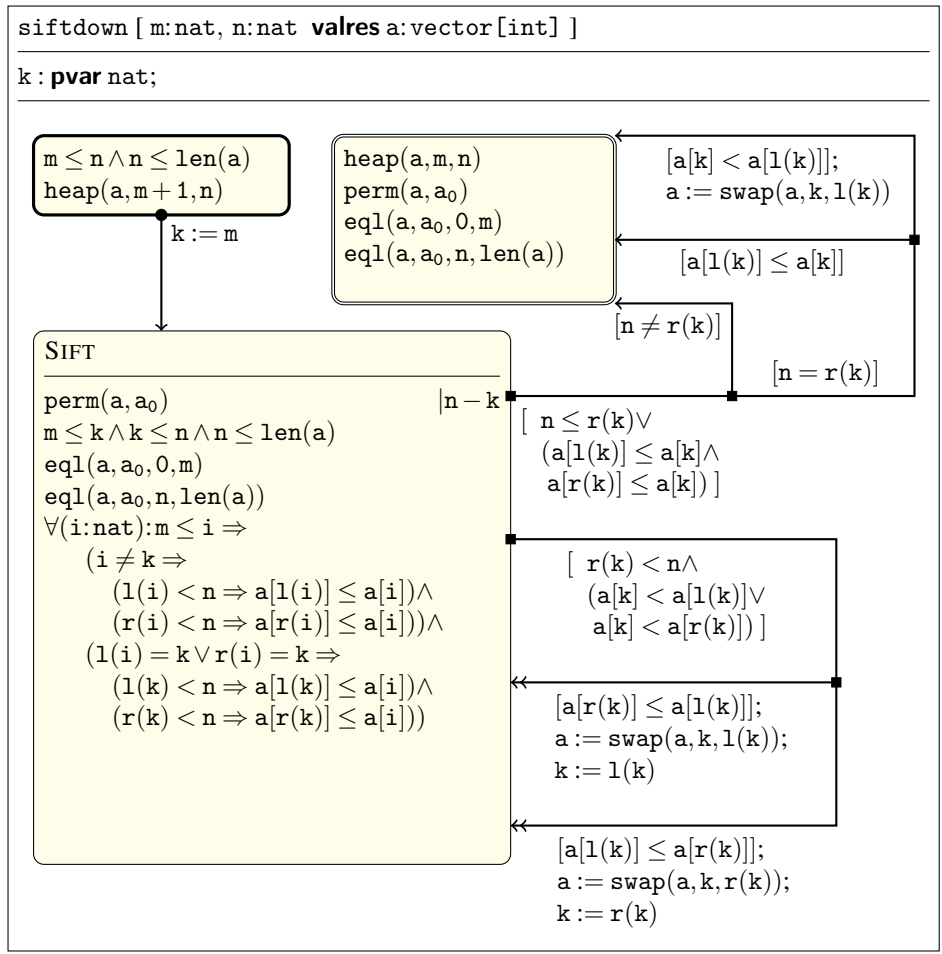

To confirm our guess that the missing corner case is the issue, we
strengthen the first disjunct of the exit guard to $\mathtt{n<r(k)}$
and re-check the program. Now, the exit transition is proved
consistent, but the liveness check for the first branch from
$\textsc{Sift}$ now fails since the case $\mathtt{n=r(k)}$ is no
longer handled. We resolve the issue by restoring the first disjunct
of the exit guard to $\mathtt{n\le r(k)}$, and handle the corner case
in a separate branch of the exit transition which swaps elements
$\mathtt{k}$ and $\mathtt{l(k)}$ if $\mathtt{a[k]<a[l(k)]}$ before
exiting to the postcondition. The updated program can be seen in
Figure \ref{fig:Siftdown-final}.  This diagram is a correct
implementation of $\mathtt{siftdown}$, and now all VCs and TCCs are
discharged automatically.

\subsection{Completing $\texttt{heapsort}$ }

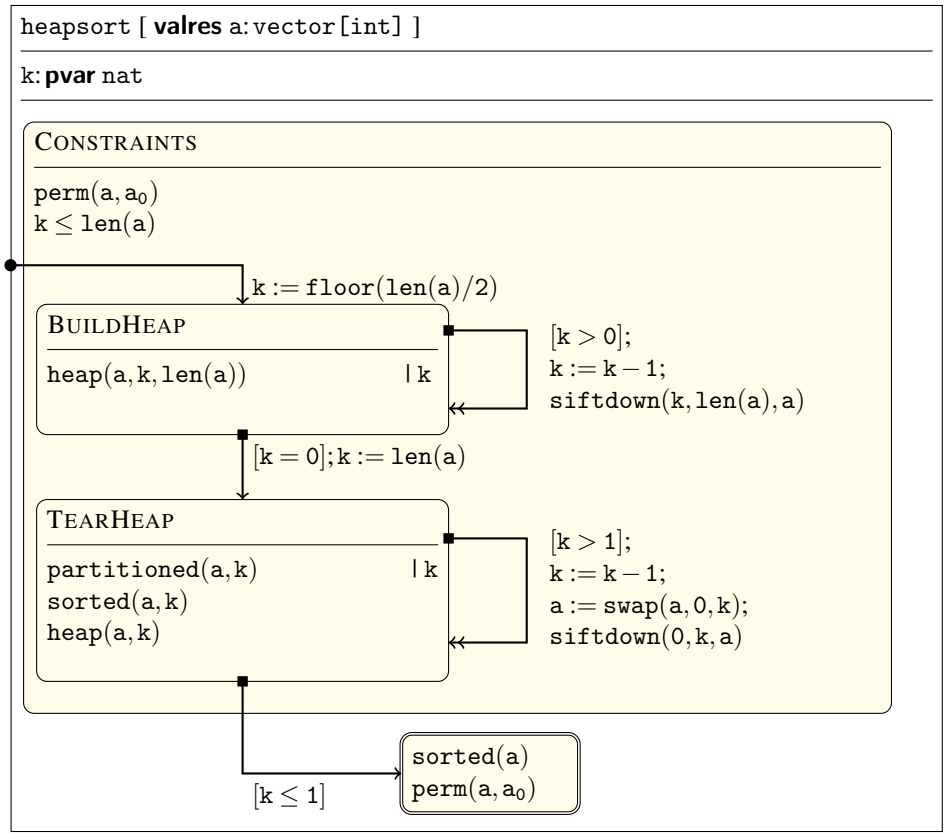
\begin{figure}
\noindent \begin{centering}
\begin{tikzpicture}
  \node[procedure] (heapsort) at (0ex,0ex) {
    \begin{minipage}{70ex}
      \procedurelabel{heapsort} [ \ibpkw{valres} \declaration{\ibpid{a} \colon \ibpid{vector[int]}} ]\\
      \procedurerule\\
      \invariant{k \colon \ibpkw{pvar}\;nat} \\
     \procedurerule\\
     \vspace{58ex}
   \end{minipage}
  };
  \node[postcondition] (Post) at (30ex,-56ex) {
      \begin{minipage}{12ex}
        \invariant{sorted(a)} \\
        \invariant{perm(a,a_0)}
      \end{minipage}
    };
  \node[situation] (Constraints) at (1ex,-9ex) {
      \begin{minipage}{65ex}
        \situationlabel{Constraints} \\
        \situationrule \\
        \invariant{perm(a,a_0)} \\
        \invariant{k \le len(a)} \\
	    \vspace{37ex}
      \end{minipage}
    };
  \node[situation] (BuildHeap) at (2ex,-23ex) {
      \begin{minipage}{30ex}
        \situationlabel{BuildHeap} \\
        \situationrule \\
        \invariant{heap(a,k,len(a))} \hfill | \invariant{k} 
        \vspace{3ex}
      \end{minipage}
    };
  \node[situation] (TearHeap) at (2ex,-38ex) {
      \begin{minipage}{30ex}
        \situationlabel{TearHeap} \\
        \situationrule \\
        \invariant{partitioned(a,k)} \hfill | \invariant{k} \\
        \invariant{sorted(a,k)} \\
        \invariant{heap(a,k)} 
        \vspace{2ex}
    \end{minipage}
    };

\draw[transition,->] (heapsort.north west) ++(0ex,-20ex) node[blackdot] {} -| node[statement] {$\mathtt{k:=floor(len(a)/2)}$} (BuildHeap.north) ;
 
\draw[transition,->] (BuildHeap.south) node[if] {} node[statement] {$\mathtt{[k=0];k:=len(a)}$} -- (TearHeap.north) ;

\draw[transition,->>] (BuildHeap.east) ++(0,3ex) node[if] {} -- ++(6ex,0) -- 
node[statement,right,anchor=west] {$
\begin{array}[t]{l}
\mathtt{[k>0];} \\
\mathtt{k:=k-1;} \\
\mathtt{siftdown(k,len(a),a)} 
\end{array}
$}   ++(0,-6ex)-- ++(-6ex,0);

\draw[transition,->>] (TearHeap.east) ++(0,4ex) node[if] {} -- ++(6ex,0) -- 
node[statement,right,anchor=west] {$
\begin{array}[t]{l}
\mathtt{[k>1];} \\
\mathtt{k:=k-1;} \\
\mathtt{a:=swap(a,0,k);} \\
\mathtt{siftdown(0,k,a)} 
\end{array}
$}   ++(0,-8ex)-- ++(-6ex,0);

\draw[transition,->] (TearHeap.south)  node[if]{} |- node[statement] {$\mathtt{[k \le 1]}$}  (Post.west);

\end{tikzpicture}
\par\end{centering}

\caption{$\mathtt{heapsort}$ with loop transitions in place\label{fig:Heapsort-complete}}
\end{figure}

Using $\mathtt{siftdown}$ to implement both missing
loop transitions, we complete the procedure $\mathtt{heapsort}$.
Figure \ref{fig:Heapsort-complete} shows the program from Figure
\ref{fig:Heapsort-initial-final-trans} extended with the loop transitions
and termination functions. 

Socos proves all termination and liveness conditions for the diagram
in Figure \ref{fig:Heapsort-complete}. It also discharges all consistency
conditions except for the $\textsc{TearHeap}$ loop transition. The
unproven condition is listed in Figure \ref{fig:Unproved-condition-Loop}.
Here, the prover has problems showing that the loop transition maintains
$\mathtt{partitioned}$. 
\begin{figure}
\begin{lstlisting}[basicstyle={\small\ttfamily},frame=lines]
    [-1]  0 <= k - 1
    [-2]  k - 1 < k
    [-3]  (heap(a_1, 0, k - 1))
    [-4]  (perm(a_1, swap(a, 0, k - 1)))
    [-5]  (eql(a_1, swap(a, 0, k - 1), 0, 0))
    [-6]  (eql(a_1, swap(a, 0, k - 1), k - 1, len(a_1)))
    [-7]  0 <= k - 1
    [-8]  k - 1 <= len(swap(a, 0, k - 1))
    [-9]  (heap(swap(a, 0, k - 1), 0 + 1, k - 1))
    [-10] k > 1
    [-11] ((k > 1 OR k <= 1))
    [-12] (perm(a, a_0))
    [-13] k <= len(a)
    [-14] (partitioned(a, k))
    [-15] (sorted(a, k))
    [-16] (heap(a, k))
      |-------
    [1]   (partitioned(a_1, k - 1))
\end{lstlisting}
\caption{Unproven condition for loop transition from $\textsc{TearHeap}$ \label{fig:Unproved-condition-Loop}}
\end{figure}
The constant $\mathtt{a\_1}$ denotes the value of $\mathtt{a}$
returned by $\mathtt{siftdown}$. The condition is hard to prove due to
the way we have defined the postcondition of $\mathtt{siftdown}$.
$\mathtt{siftdown}$ manipulates the leftmost portion of the array, and
the properties of $\mathtt{perm}$ given to the automatic prover cannot
be used to infer that $\mathtt{partitioned}$ is maintained throughout
the procedure call. Proving the condition actually requires two
non-trivial properties: 1) the root of a max-heap is the maximal
element; and 2) if $\mathtt{partitioned}$ holds for an index and an
array, it also holds for a permutation of the array where the portion
to the right of the index is unchanged.  One alternative is to start
proving this condition directly in PVS.  However, it is better to
first make properties (1) and (2) explicit in the program by adding
assert statements to the loop transition:
\begin{quote}
$\begin{array}[t]{l}
\mathtt{\color{gray}{[k>1];}}\\
\mathtt{\color{gray}{k:=k-1\,;}\,\color{black}{\{\forall(i:index(a)):i\le k\Rightarrow a[i]\le a[0]\};}} \\
\mathtt{\color{gray}{a:=swap(a,0,k)\,;}\,\color{black}{\{partitioned(a,k)\}\,;}} \\
\mathtt{\color{gray}{siftdown(0,k,a)}}
\end{array}$
\end{quote}
Re-checking, we are left with two simpler conditions:
the first assertion above, and the condition from Figure
\ref{fig:Unproved-condition-Loop} but with the above assertions as
additional antecedents. The second assertion is discharged
automatically. The first assertion can be proved with a
straightforward induction proof. Proving that
$\mathtt{partitioned(a\_1,k-1)}$ is a consequence of
$\mathtt{partitioned(swap(a,0,k-1),k-1)}$ and the antecedents in
Figure \ref{fig:Unproved-condition-Loop} is much more involved,
requiring reasoning in terms of the definition of permutation. To
finish the verification, we prove the lemmas $\texttt{heap\_max}$
and $\texttt{perm\_partitioned}$ in the background theory:
\begin{quote}
$\begin{array}[t]{l}
\texttt{heap\_max}\mathtt{:}\ \textsf{\textbf{lemma}}\\
\quad\mathtt{\forall(k\colon nat):heap(a,0,k)\Rightarrow(\forall(i\colon nat):i<k\Rightarrow a(i)\le a(0))}\\
\\
\texttt{perm\_partitioned}\mathtt{:}\ \textsf{\textbf{lemma}}\\
\quad\mathtt{\forall(a,b,(k:upto(len(a)))):\begin{array}[t]{@{}l}
\mathtt{perm(a,b)\land partitioned(a,k)\land eql(a,b,k,len(a))}\\
\mathtt{\Rightarrow partitioned(b,k)}
\end{array}}
\end{array}$
\end{quote}
With the help of these additional lemmas, the condition can be 
discharged automatically.

\section{Conclusion\label{sec:Conclusion}}

In this paper, we have described the Socos environment and shown how
it combines specification, implementation and verification of
invariant-based programs into a single workflow. We demonstrated the
use of Socos in construction of a correct invariant-based
implementation of heapsort.  The full verification workflow comprised
three sequential stages.  First background theories for arrays,
sorting and permutations were built in PVS. Secondly, the situation
structure, consisting of the specifications and internal loop
invariants, was defined. Thirdly, the transitions were added and
verified consistent with the situations. The result is a PVS checked
proof of consistency, liveness and termination of the invariant
diagram.

The $\texttt{endgame}$ strategy, which relies on the SMT solver Yices,
automatically discharges most of the simple verification conditions.
When $\texttt{endgame}$ is unable to discharge a true condition,
we have the following options to proceed: 
\begin{itemize}
\item Prove the condition interactively in PVS; however, since such
  proofs are closely coupled to the implementation, they are sensitive
  to changes in the code and/or specification.
\item Add an assume statement to achieve consistency at the cost of
  liveness;  this is a valid alternative if full verification is not
  required because we are satisfied with, e.g., testing the parts that
  could not be automatically verified.
\item Add an assert statement to isolate a specific difficult
  condition on which the proof depends; this condition can then be
  handled using one of the other alternatives.
\item Add a helper lemma to the background theory, prove it, and ask
  $\texttt{endgame}$ to apply it automatically.
\end{itemize}
The case study presented in Section \ref{sec:Example:Heapsort} used
background theories extensively. The properties introduced in the
theories are reasonably general, and could be reused in other
verification contexts. The actual application of the lemmas to verify
individual transitions was completely automatic. In our experience,
extending the default strategy with additional lemmas should be done
judiciously, since they increase the size of the verification problem.
Adding too many lemmas may cause the SMT solver to hit time or memory
constraints. When this issue develops, the different parts of the
program that depend on separate background theories must be identified
and verified separately.  In general, our experience has been that
careful formulation of the background theory and the situation
structure of the program are the key elements to successfully
integrating programming and proving.

\begin{small}
 \bibliographystyle{eptcs}
 \bibliography{thedu11}
\end{small}

\end{document}